\documentclass[jmanuscript=article,etalmode=truncate,maxauthors=0,journal=apchd5]{achemso}

\usepackage{chemformula} % Formula subscripts using \ch{}
\usepackage[T1]{fontenc} % Use modern font encodings
\usepackage[colorlinks, linkcolor= blue, citecolor = blue, urlcolor=blue]{hyperref}
\usepackage{soul}
\usepackage{graphicx}
\usepackage{amsmath}
\usepackage{amssymb}
\usepackage{hyperref}

\setkeys{acs}{etalmode=truncate}
\setkeys{acs}{maxauthors = 0}

\newcommand\beq{\begin{equation}}
	\newcommand\eeq{\end{equation}}

\author{Massimo Moccia}
\affiliation{Fields \& Waves Lab, Department of Engineering, University of Sannio, Benevento, I-82100, Italy}
\author{Giuseppe Castaldi}
\affiliation{Fields \& Waves Lab, Department of Engineering, University of Sannio, Benevento, I-82100, Italy}
\author{Andrea Al\`u}
\affiliation{Photonics Initiative, Advanced Science Research Center, City University of New York, New York, NY 10031, USA}
\affiliation{Physics Program, Graduate Center, City University of New York, New York, NY 10016, USA}
\author{Vincenzo Galdi} \email{vgaldi@unisannio.it}
\affiliation{Fields \& Waves Lab, Department of Engineering, University of Sannio, Benevento, I-82100, Italy}
\email{vgaldi@unisannio.it}

\title[Ghost Line Waves]
  {Ghost Line Waves}

\abbreviations{IR,NMR,UV}

\begin{document}


\begin{abstract}
Time-harmonic electromagnetic plane waves  in anisotropic media can exhibit complex-valued wavevectors (with nonzero real and imaginary parts) even in the absence of material dissipation. These peculiar modes, usually referred to as ``ghost waves'', hybridize the typical traits of conventional propagating and evanescent waves, displaying both phase accumulation and purely reactive exponential decay away from the direction of power flow. Their existence has been predicted in several scenarios, and has been recently observed experimentally in the form of surface phonon polaritons with complex-valued out-of-plane wavevectors 
propagating
at the interface between air and  a natural uniaxial crystal with slanted optical axis. Here, we 
demonstrate that ghost waves can arise also in lower-dimensional flat-optics scenarios, which are becoming increasingly relevant 
in the context of metasurfaces and in the field of polaritonics. Specifically, we show that planar  junctions between isotropic  and anisotropic metasurfaces can support ``ghost line waves'' that propagate unattenuated along the line interface, exhibiting  phase oscillations combined with evanescent decay both in the plane of the metasurface (away from the interface) and out-of-plane n the surrounding  medium. Our theoretical results, validated by finite-element numerical simulations,  demonstrate a novel  form of polaritonic waves with highly confined features, which may provide new
opportunities for the control of light at the nanoscale, and
may find potential applications
in a variety of scenarios, including integrated waveguides, nonlinear optics, optical sensing  and  sub-diffraction imaging. 
\end{abstract}

%%%%%%%%%%%%%%%%%%%%%%%%%%%%%%%%%%%%%%%%%%
\section*{Introduction}
%%%%%%%%%%%%%%%%%%%%%%%%%%%%%%%%%%%%%%%%%%
In a transparent medium, the most common type of time-harmonic electromagnetic wave 
is the propagating (or ``uniform'') plane wave, characterized by a real-valued wavevector, i.e., unattenuated oscillations and active power flow along the propagation direction \cite{Harrington:2001th}. 
In the case of  total reflection at an interface, plane waves can become  evanescent (``nonuniform''), characterized by wavevectors with either real-valued or purely imaginary components, resulting in phase propagation and exponential attenuation (with purely reactive power flow) along two orthogonal directions \cite{Harrington:2001th}. Recent studies have theoretically predicted the existence of a new type of plane waves in lossless biaxial anisotropic materials, characterized by a {\em complex-valued} wavevector  (with nonzero real and imaginary parts) \cite{Narimanov:2018dw,Narimanov:2019gr}. These modes, named ``ghost waves'' (GWs) in analogy with ghost orbits in quantum mechanics \cite{Kus:1993pp}, combine some typical traits of propagating and evanescent waves, exhibiting both phase accumulation and purely reactive power flow along the same direction (see also Refs. \citenum{Mackay:2019eg,Waseer:2019nu} for general theoretical considerations on nonuniform plane waves).

More recent studies have focused on the surface-wave (SW) realization of GWs. In particular, ghost surface polaritons (GSP) have been theoretically predicted in ionic-crystal (ZnS/SiO$_2$) metamaterials \cite{Zhou:2018gs} and antiferromagnets \cite{Song:2022gs}, and have been experimentally observed in a bulk calcite crystal \cite{Ma:2021gh}. Even in an idealized lossless scenario, these wave modes are characterized by a complex-valued out-of-plane wavevector, thereby combining the traits of conventional bulk and surface polaritons, with both propagating (phase accumulation) and evanescent (exponential decay) behavior inside the crystal. Their in-plane propagation on the crystal surface is instead characterized by hyperbolic dispersion, giving rise to  long-range, directional and diffractionless polariton propagation \cite{Ma:2021gh}.

It is important to distinguish between GWs and {\em leaky} waves, as both exhibit complex-valued wavevectors but have distinct properties. Leaky waves are complex eigenmodes that can be supported by open waveguiding structures and are compatible with radiation in terms of their momentum; they can effectively model radiative phenomena from such systems \cite{Monticone:2015lw}. On the other hand, GWs are complex bound modes that rely on broken symmetries for their realization, with the necessity to operating with biaxial and slanted optical axes with respect to an interface.

Far from being physical oddities of mere academic concern, GWs and GSPs 
exhibit several unusual features of great interest, including the possibility to
exponentially enhance and modulate incident evanescent waves via resonant coupling mechanisms \cite{Narimanov:2019gr}, to attain exact frequency degeneracies in guided modes \cite{Debnath:2021gi},
and to control nanoscale light propagation via the optical-axis orientation \cite{Ma:2021gh}. These may find important applications in a variety of scenarios, ranging from integrated waveguides and nonlinear optics \cite{Narimanov:2019gr,Debnath:2021gi} in optical sensing \cite{Khan:2022gs} and  sub-diffraction imaging \cite{Ma:2021gh}.

Expanding the aforementioned features and their associated applications to flat-optics platforms becomes a fascinating development.
As such, we demonstrate here the possibility of exciting GWs in  planar junctions of metasurfaces. In analogy with the higher-dimensional concept of surface polaritons, here we consider the recently introduced concept of {\em line waves} (LWs), which can propagate along metasurface junctions characterized by oppositely signed surface reactances \cite{Horsley:2014od,Bisharat:2017ge} or resistances \cite{Moccia:2020lw}. Conventional LWs at isotropic metasurface junctions exhibit unattenuated propagation along the interface and evanescent (monotonic) decay both out-of plane (away from the metasurfaces, as conventional SWs) and in-plane (away from the interface), thereby effectively channeling the power flow along a one-dimensional waveguide (hence the name) \cite{Horsley:2014od,Bisharat:2017ge}, with tremendous benefits for guiding in an ultra-subdiffractive way light and polaritons.

Here we show that, in the presence of (lossless) anisotropy, a new kind of LW can be supported exhibiting GW-type behavior (i.e., phase oscillations combined with evanescent decay) both in-plane (along the metasurface) and out-of-plane (in the surrounding  medium). These ghost LWs (GLWs) can be viewed as the one-dimensional version of GWs \cite{Narimanov:2018dw,Narimanov:2019gr} and GSPs \cite{Zhou:2018gs,Song:2022gs,Ma:2021gh}, and they share several intriguing features with their higher-dimensional equivalents, opening up
new perspectives in interface nano-optics for the advanced control of surface polaritons in  deeply sub-diffractive regimes \cite{Zhang:2021in,Wu:2022mp}.

%%%%%%%%%%%%%%%%%%%%%%%%%%%%%%%%%%%%%%%%%%
\section*{Results and discussion}
%%%%%%%%%%%%%%%%%%%%%%%%%%%%%%%%%%%%%%%%%%

%############################################################
%                Figure1
%
\begin{figure}[ht]
	\centering
	\includegraphics[width=.8\linewidth]{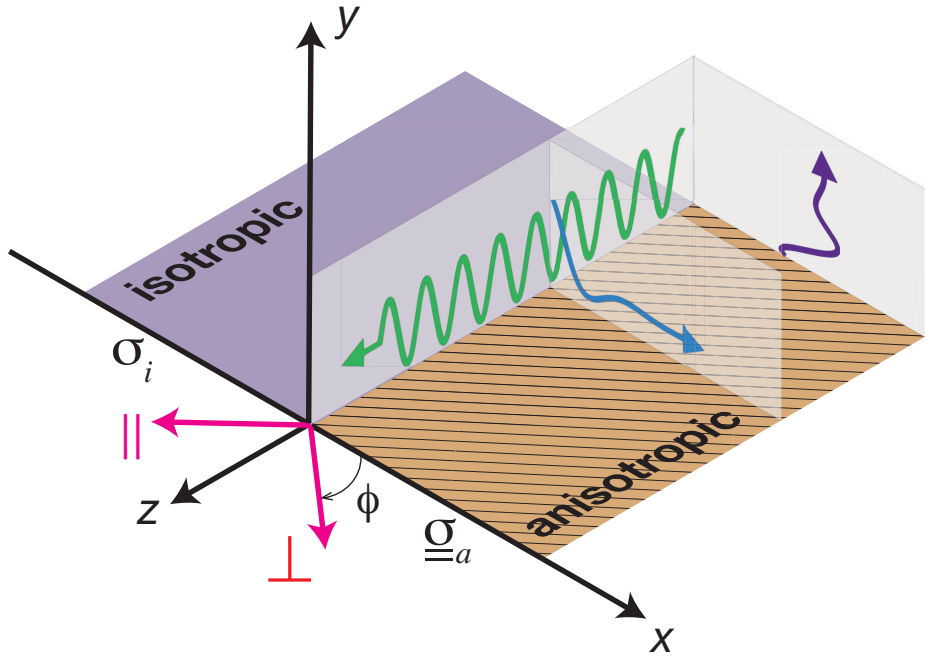}
	\caption{Problem schematic: A planar junction between an isotropic metasurface and an anisotropic, reciprocal one with slanted optical axis. The wavy arrows depict a GLW that propagates unattenuated along the interface, and exhibits an oscillatory decay both in-plane and out-of-plane in the anisotropic halfplane ($x>0$). In the isotropic region ($x<0$), a monotonic decay (not shown) similar to conventional LWs is anticipated. Also shown are the absolute ($x,y,z$) and principal ($\perp, y, \parallel$) reference systems.}
	\label{Figure1}
\end{figure}
%############################################################

Referring to Figure \ref{Figure1} for a conceptual illustration, we consider a  junction in the $x-z$ plane between an isotropic metasurface (characterized by a surface conductivity $\sigma_i$) and an anisotropic, reciprocal one (characterized by the parallel and orthogonal surface conductivities $\sigma_{\parallel}$ and $\sigma_{\perp}$, respectively) with an optical axis rotated by an angle $\phi$. Alternatively, the anisotropic metasurface can be described, in the $(x,z)$ in-plane reference system, by a symmetric tensor ${\underline {\underline \sigma}}_a$ with components
\begin{subequations}
	\begin{eqnarray}
		{\sigma_{xx}} &=& {\sigma _{\perp}}{\cos ^2}\phi  + {\sigma_{\parallel}}{\sin ^2}\phi,\\
		{\sigma_{zz}} &=& {\sigma_{\parallel}}{\cos ^2}\phi  + {\sigma_{\perp}}{\sin ^2}\phi,\\
		{\sigma_{xz}} &=& {\sigma_{zx}}=\left( {{\sigma_{\perp}} - {\sigma_{\parallel}}} \right)\cos \phi \sin \phi.
	\end{eqnarray}
\end{subequations}
This is a commonly utilized model for nanostructured  metasurfaces \cite{Gomez-Diaz:2016fo}  as well as for natural van der Waals crystals \cite{Ma:2018ip}.
For now, we assume that the metasurfaces are purely reactive (i.e., purely imaginary surface conductivity values) and immersed in vacuum, so that the entire scenario is ideally lossless; this is done to ascertain the genuine character of GW-type solutions, since the presence of losses would naturally induce complex-valued wavevectors. As conceptually sketched in Figure \ref{Figure1}, we are interested in finding GLW solutions that propagate unattenuated along the interface $x=0$ and, in the anisotropic region ($x>0$), decay with an oscillating character both in-plane and out-of-plane. Conversely, in the isotropic region $x<0$, we anticipate a monotonic decay (not shown) similar to conventional LWs.

The boundary-value problem in Figure \ref{Figure1} admits an analytical (Maliuzhinets-type) solution for isotropic junctions \cite{Kong:2019at},
whereas the anisotropic case of interest here  is insofar largely unexplored. Previous studies \cite{Sokolik:2021pm} are restricted to the investigation of ``edge modes'' in the particular case of a truncated anisotropic metasurface (corresponding to $\sigma_i=0$ in Figure \ref{Figure1}) via a Wiener-Hopf  approach with a Fetter-type approximation. Here, in order to consider the more general scenario in Figure \ref{Figure1}, we follow a numerical approach based on finite-element simulations.
However, some useful physical insight can be obtained via a preliminary analytical study of the SWs supported by the anisotropic half assumed as of infinite extent (for the well-known isotropic case, see, e.g., Ref. \citenum{Collin:1969at}). Accordingly, we are interested in time-harmonic modal solutions of the form $\propto\exp \left[ i\left(k_x x + k_y y + k_z z -\omega t\right)\right]$.  These modes are generally hybrid, and their dispersion equation can be written as \cite{Bilow:2003gw,Patel:2013ma}
\begin{subequations}
	\beq
	2\eta\left[
	{\sigma_{zz}} \left(k_x^2+k_y^2\right)+{\sigma_{xx}} \left(k_y^2+k_z^2\right)- 2{\sigma_{xz}{k_x}{k_z}}
	\right]
	+kk_y\left[\eta^2\left(\sigma_{xx}\sigma_{zz}-\sigma_{xz}^2\right)+4\right]=0,
	\label{eq:DE1}
	\eeq
	with the constraint	
	\beq
	k_x^2 + k_y^2 + k_z^2 = k^2,\quad \mbox{Im}\left(k_y \right) \ge 0,
	\label{eq:PW}
	\eeq
	\label{eq:DE}
\end{subequations}
where $\eta$ and $k=\omega/c=2\pi/\lambda$ are the vacuum intrinsic impedance and wavenumber, respectively (with $c$ and $\lambda$ being the corresponding wavespeed and wavelength, respectively), and the branch-cut is chosen so as to ensure the physically feasible out-of-plane decay away from the metasurface. From eqs \ref{eq:DE}, we can calculate the iso-frequency contours, by fixing $\omega$, varying one (real-valued) wavevector component (say $k_z$), and deriving the remaining two ($k_x, k_y$) by solving analytically a quartic equation (see Supporting Information for details). Of the resulting four modal branches, only some (typically two) exhibit the proper branch-cut $\mbox{Im}(k_y)\ge 0$. In what follows, we focus on these proper solutions and, without loss of generality, we consider the forward-propagating case ($k_z>0$) and  the angular range $0<\phi<90^{\circ}$; outside these ranges, the solutions follow by symmetry considerations (see Supporting Information for details). Moreover, for notational convenience, we introduce the (dimensionless) normalized conductivities \cite{Ma:2018ip}
\beq
\alpha_{\nu}=2\pi\eta\sigma_{\nu},\quad \nu=i, \parallel, \perp.
\eeq

%############################################################
%                Figure2
%
\begin{figure}
	\centering
	\includegraphics[width=\linewidth]{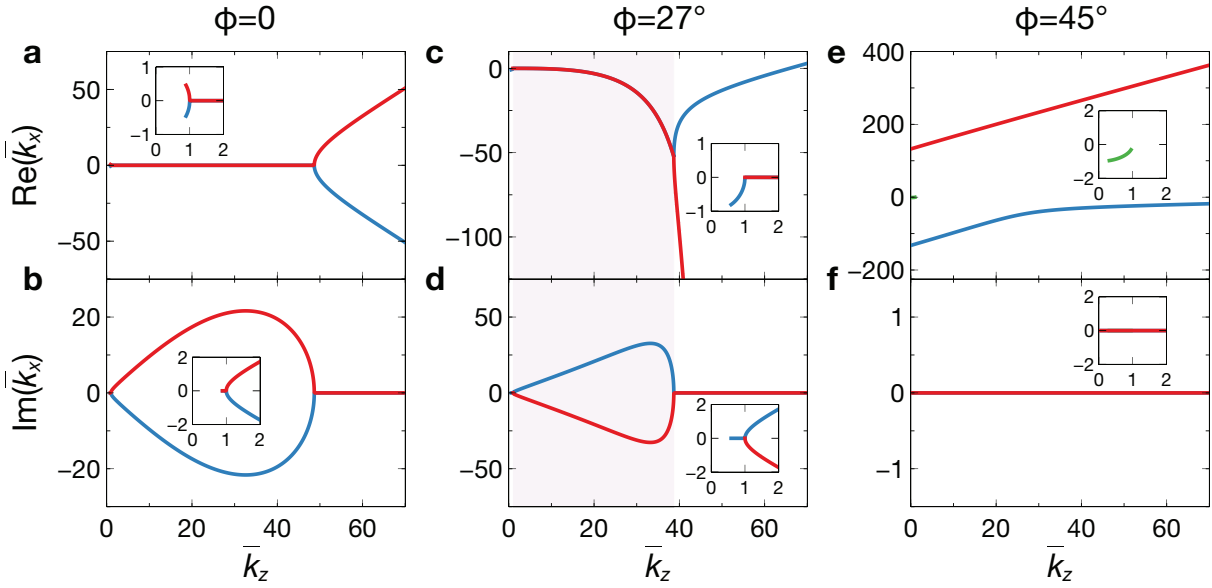}
	\caption{Representative iso-frequency contours for hyperbolic anisotropy. (a,b) Real and imaginary parts of normalized wavenumbers ${\bar k}_x=k_x/k$ as a function of $\bar{k}_z=k_z/k$, for a hyperbolic metasurface 
		with $\alpha_{\perp}=-i0.07$, $\alpha_{\parallel}=i0.26$,  and $\phi=0$. Only the proper modal branches with $\mbox{Im}(k_y)\ge 0$ are shown, for $k_z>0$, with the insets displaying some magnified views around ${\bar k}_z=1$. (c,d) Same as panels (a,b), respectively, but for $\phi=27^\circ$. The pink-shaded region delimits the GW range.
		(e,f) Same as panels (a,b), respectively, but for $\phi=45^\circ$.}
	\label{Figure2}
\end{figure}
%############################################################

Akin to their higher-dimensional counterparts, GLWs originate from broken symmetries. For a representative illustration,
we consider a hyperbolic metasurface with normalized conductivities $\alpha_{\perp}=-i0.07$ and $\alpha_{\parallel}=i0.26$, which are compatible with the values reported in the literature\cite{Ma:2018ip} for a thin layer of $\alpha$-MoO$_3$. 
Figure \ref{Figure2} shows the corresponding iso-frequency contours (in terms of normalized wavenumbers ${\bar k}_{x,z}=k_{x,z}/k$) for three representative values of the rotation angle $\phi$.
Specifically, Figures \ref{Figure2}a,b pertain to the well-known \cite{Gomez-Diaz:2016fo} symmetric scenario in the absence of rotation (i.e., $\phi=0$), for which we observe the expected presence of two propagating modes at small (see insets) and large wavenumbers; these solutions are conventional SWs whose fields  extend in-plane and exponentially decay out-of-plane. The two propagating regions are separated by a bandgap corresponding to evanescent modes (i.e., purely imaginary $k_x$ and $k_y$) that decay exponentially both in-plane (along $x$) and out-of-plane. As it will be clearer hereafter, this spectral region is compatible with the propagation of conventional LWs in junction-type configurations as in Figure \ref{Figure1}.
However, by rotating the optical axis, as for instance in Figures \ref{Figure2}c,d ($\phi=27^\circ$), a symmetry-breaking deformation of the iso-frequency contours occurs, resulting in a nonzero real part in $k_x$ (and thus, from eq \ref{eq:PW}, in $k_y$) over an otherwise evanescent region
 $k<k_z\lesssim39k$ (highlighted with pink shading). These modes exhibit the aforementioned GW hallmarks (phase accumulation and exponential decay) both in-plane and out-of-plane. It is interesting to note that, unlike the volumetric scenario \cite{Narimanov:2018dw,Narimanov:2019gr}, now the GW-type behavior is also induced in a material (vacuum, in our example) that is not anisotropic; while this can be expected for lossy, isotropic metasurfaces, it remains a remarkable feature in our lossless scenario, which may provide new possibilities for applications.
We observe that the two GW branches are characterized by complex-conjugate wavenumbers $k_{x1}=k_{x2}^*$ (and, from eq \ref{eq:PW}, $k_{y1}=-k_{y2}^*$). Clearly, for a metasurface of infinite extent, only the decaying branch [$\mbox{Im}(k_x)\ge 0$] is physically feasible but, in the presence of spatial truncations, the other branch could resonantly couple with an incident evanescent field, and amplify it through the anisotropic section, along the lines of what predicted for conventional GWs \cite{Narimanov:2019gr}.
Beyond this GW spectral region (i.e., $k_z\gtrsim39k$), we recover conventional SWs as for the unrotated scenario.

Larger rotation angles (e.g., $\phi={45}^{\circ}$) may cause even stronger deformations in the iso-frequency contours (see Figures \ref{Figure2}e,f), such as the transition of propagating modes from improper to proper character (and viceversa), leading to the vanishing of evanescent and GW regions, with the presence of propagating SWs only.

%############################################################
%                Figure3
%
\begin{figure}
	\centering
	\includegraphics[width=.6\linewidth]{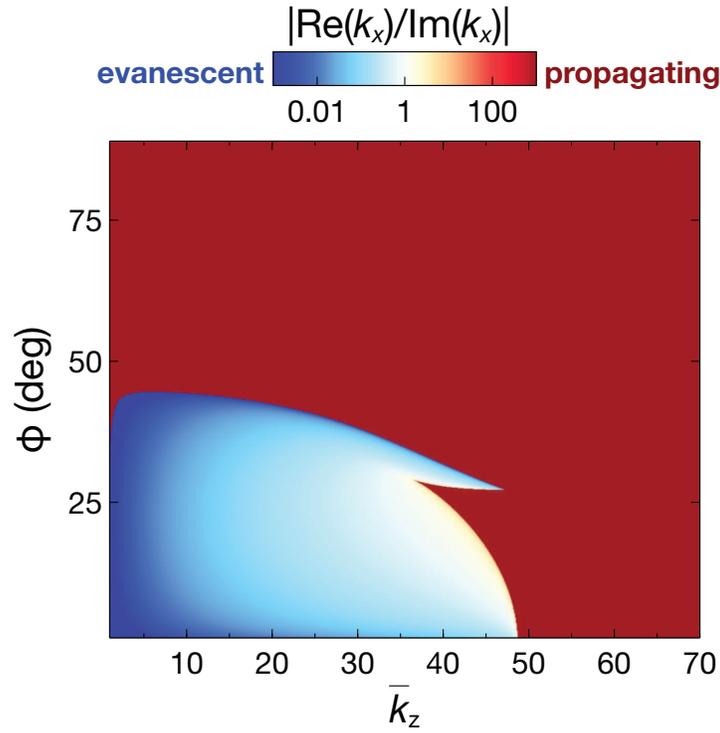}
	\caption{Parametric study for hyperbolic anisotropy. False-color-scale map of $\left|\mbox{Re}\left(k_x\right)/\mbox{Im}\left(k_x\right)\right|$, as a function of rotation angle $\phi$ and normalized wavenumber ${\bar k}_z$, for a hyperbolic metasurface with parameters as in Figure \ref{Figure2} ($\alpha_{\perp}=-i0.07$, $\alpha_{\parallel}=i0.26$). Due to symmetry, only the ranges $k_z>0$ and  $0<\phi<90^{\circ}$ are displayed.}
	\label{Figure3}
\end{figure}
%############################################################

For a broader illustration, Figure \ref{Figure3} shows the parameter $\left|\mbox{Re}\left(k_x\right)/\mbox{Im}\left(k_x\right)\right|$ as a function of $k_z$ and the rotation angle $\phi$. This parameter quantifies the oscillatory vs. damped character, with the asymptotic limits $0$ and $\infty$ corresponding to purely evanescent and propagating modes, respectively, and finite values indicating the intermediate GW regime. We find that GW-type modes can exist within a limited interval of rotation angles and wavenumbers, with the allowed spectral region progressively shrinking as $\phi$ approaches a critical value of $\sim44^\circ$.

Qualitatively similar results are observed for different parameter choices, which mainly affect the rotation-angle and wavenumber ranges where GW-type modes can exist.

Overall, the above results indicate that, in principle, generic hyperbolic metasurfaces  with slanted optical axes can support GW-type modal solutions within certain spectral ranges. 
The question now arises on what physical mechanism could be exploited to induce the desired (real-valued) in-plane wavenumber $k_z$. A possible answer is provided by  the metasurface junction in Figure \ref{Figure1}. The basic idea, in analogy with the GSP scenario \cite{Zhou:2018gs,Song:2022gs,Ma:2021gh}, is to tailor the surface reactance of the isotropic half so that a LW can propagate along the metasurface junction with 
a real-valued propagation constant $k_z^{(LW)}$. From analogies with the isotropic-junction scenario \cite{Horsley:2014od,Bisharat:2017ge,Kong:2019at}, we expect the  field behavior at the interface to possibly exhibit a singularity, and therefore be quite different from conventional SWs. Nevertheless, at a sufficient distance (along $x$) from the interface, it is reasonable to assume that the field is approximately ruled by the infinite-metasurface dispersion relationship, with $k_z\approx k_z^{(LW)}$.

%############################################################
%                Figure4
%
\begin{figure}
	\centering
	\includegraphics[width=.6\linewidth]{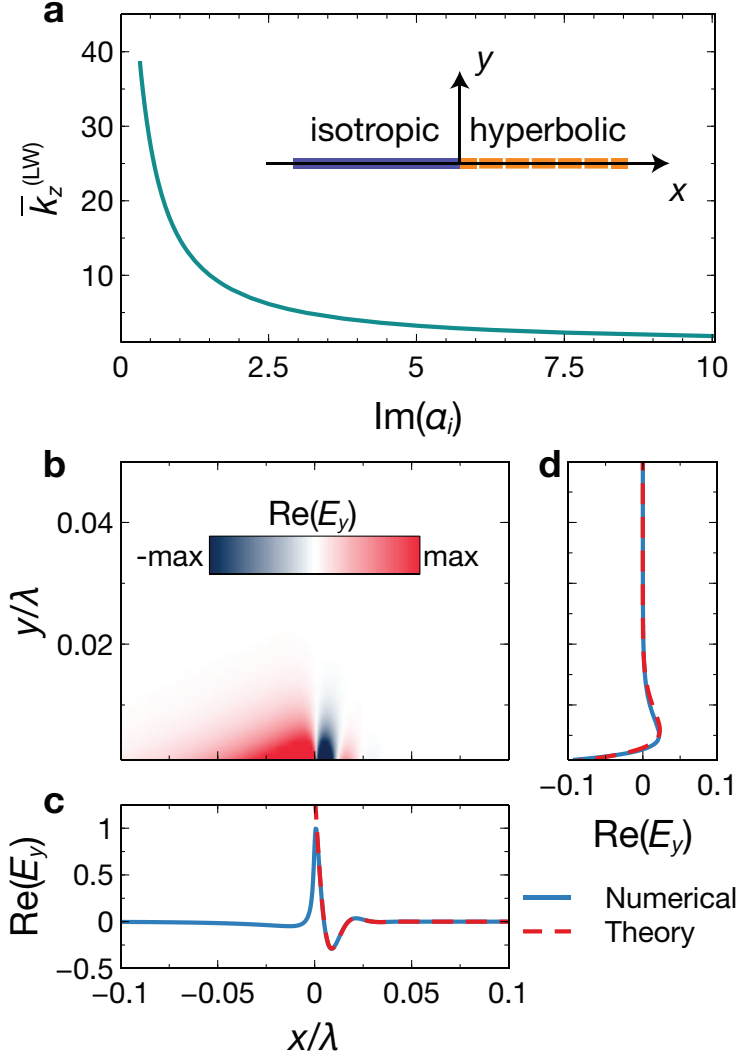}
	\caption{Eigenmode analysis for isotropic-hyperbolic junction. (a) Numerically computed LW effective modal index for a metasurface junction as in Figure \ref{Figure1} (see also inset), as a function of $\mbox{Im}\left(\alpha_i\right)$, for a hyperbolic half with  parameters as in Figures \ref{Figure2}c,d ($\alpha_{\perp}=-i0.07$, $\alpha_{\parallel}=i0.26$, $\phi=27^\circ$). (b) Numerically computed field-map [$\mbox{Re}(E_y)$] for $\alpha_i=i0.344$ (i.e., ${\bar k}_z^{(LW)}=37$). (c,d) Corresponding cuts (blue-solid) at $y=0.001\lambda$ and $x=0.0054\lambda$, respectively, compared with theoretical predictions (red-dashed) from eqs \ref{eq:DE} with $k_z=k_z^{(LW)}$, yielding ${\bar k}_x=-39.88 +i26.38$ and ${\bar k}_y=20.33+i51.73$. Fields are normalized with respect to maximum value.}
	\label{Figure4}
\end{figure}
%############################################################

To verify the above reasoning and assumptions, we carry out a numerical eigenmode analysis (see Supporting Information for details) of this structure. 
Assuming a hyperbolic half with parameters as in Figures \ref{Figure2}c,d, we show in Figure \ref{Figure4}a the numerically computed LW effective modal index ${\bar k}_z^{(LW)}=k_z^{(LW)}/k$ as a function of the (imaginary) normalized conductivity $\alpha_i$ of the isotropic half. 
Indeed, by varying $\alpha_i$ within the inductive range [$\mbox{Im}\left(\alpha_i\right)>0$], the junction
supports a LW mode with real-valued $k_z$; notably, the range of achievable $k_z$ values aligns well with our theoretical prediction of the GW spectral range (as shown in Figures \ref{Figure2}c,d).

The above results provide a key connection between  our theoretical study in Figure \ref{Figure2} and the actual physical structure in Figure \ref{Figure1}, showing that this latter can support GLW-type modes by suitably tailoring the surface conductivity of the isotropic half. Interestingly, the required conductivity values are in line with those attainable from two-dimensional materials such as graphene at THz frequencies \cite{Vakil:2011to}. We highlight that, given the hyperbolic character of the anisotropic half, there is no clear-cut choice of a ``complementary'' isotropic half in order to support a LW.
For this specific example, an inductive isotropic metasurface is needed,  but different parameter choices may require  a capacitive one
[i.e., $\mbox{Im}\left(\alpha_i\right)<0$].

For a specific value of $\alpha_i$ (corresponding to ${k}_z^{(LW)}=37k$), Figure \ref{Figure4}b shows a field-map of the numerically computed eigenmode [$\mbox{Re}(E_y)$; see Supporting Information for details on the numerical simulations and the complete field maps], from which the typical GW hallmarks (oblique wavefronts) are clearly observed in the anisotropic half. In the isotropic half, instead, the in-plane and out-of-plane attenuation is monotonic, as in conventional LWs \cite{Horsley:2014od,Bisharat:2017ge}.
For a more quantitative assessment, Figures \ref{Figure4}c and \ref{Figure4}d show two representative cuts (at fixed $y$ and $x$, respectively), from which we observe the expected oscillatory damping, both in-plane and out-of-plane, in excellent agreement with our theoretical predictions. As previously mentioned, these latter are complex exponentials, with propagation constants $k_x$ and $k_y$ obtained from eqs \ref{eq:DE} by replacing $k_z$ with the numerically computed LW propagation constant $k_z^{(LW)}$. Therefore, these modes can be viewed as the one-dimensional version of GSPs and conventional GWs; it can be verified that the oscillatory damping does not imply actual active power flow away from the junction (see Supporting Information for details).

As anticipated, in the absence of rotation ($\phi=0$, i.e., symmetric case), the modal solutions may take the form of conventional LWs, which decay monotonically for both positive and negative values of $x$, as well as along $y$ (see Supporting Information for a representative example). On the other hand, for scenarios where an extended behavior  is expected (such as for $\phi={45}^{\circ}$, or for $\phi={27}^{\circ}$ within the spectral range $k_z>40k$), a two-dimensional eigenmode analysis is not as meaningful.

The above results validate our theoretical reasoning and assumptions, indicating
that it is actually possible to design a planar junction that supports a GLW, whose properties can be  tailored and controlled by varying the optical-axis slant and/or the conductivity of the isotropic half.

%############################################################
%                Figure5
%
\begin{figure}
	\centering
	\includegraphics[width=1\linewidth]{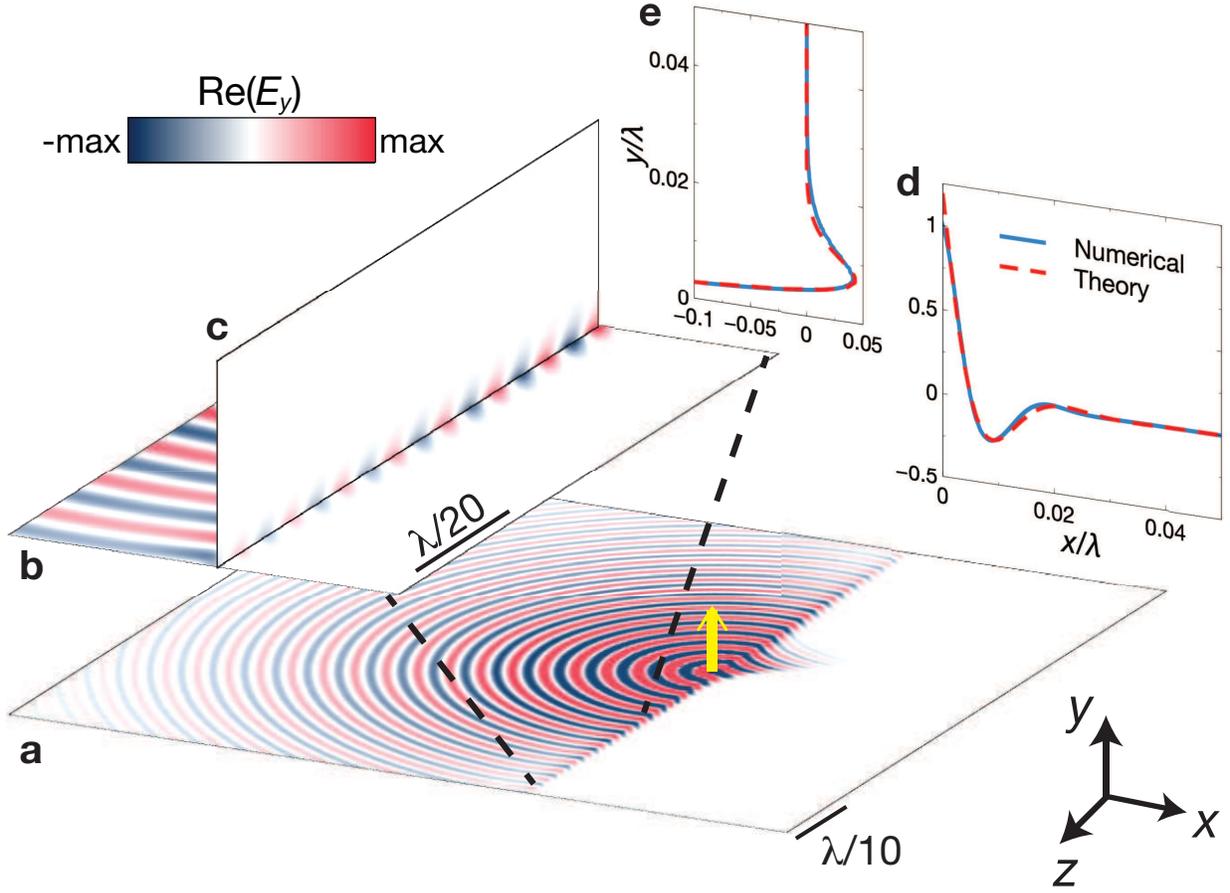}
	\caption{Dipole excitation and losses for isotropic-hyperbolic junction. (a) Numerically computed in-plane field map $[\mbox{Re}(E_y)]$ at $y=0.003\lambda$, for a configuration with parameters as in Figure \ref{Figure3}b but with losses ($\alpha_{\perp}=0.0035-i0.07$, $\alpha_{\parallel}=0.0013+i0.26$, $\phi=27^\circ$, $\alpha_i=0.0034+i0.344$), excited by a $y$-directed elementary electric dipole located at $x=z=0$, $y=0.01\lambda$ (schematized by a thick yellow arrow).
		(b,c) Magnified views in-plane ($y=0.003\lambda$) and out-of-plane ($x=0.0057\lambda$), respectively. (d,e) Corresponding cuts (blue-solid) at $z=0.3\lambda$, with $y=0.003\lambda$ and $x=0.0057\lambda$, respectively, compared with theoretical predictions (red-dashed). Fields are normalized with respect to maximum value.}
	\label{Figure5}
\end{figure}
%############################################################

As a last step, to verify the actual excitability of the above eigenmode solutions, we carry out three-dimensional numerical simulations in the presence of a physical excitation (see Supporting Information for details). Moreover, we take into account the possible presence of losses, assuming for the parameter configuration in Figure \ref{Figure4} a loss-tangent of 0.05 for the hyperbolic half (compatible with the values reported \cite{Yu:2022uq} for $\alpha$-MoO$_3$) and 0.01 for the isotropic one (compatible with typical values of graphene \cite{Vakil:2011to}).

Figure \ref{Figure5}a shows the in-plane field distribution for a dipole-excited configuration. As expected, this source also excites (isotropic and anisotropic) conventional SWs that are naturally supported by the two metasurface halves. In principle, these SWs could be de-emphasized via suitable optimization of the parameters and sources, but this was not a concern here, since their presence does not overshadow the GLW phenomenon of interest. 
Indeed, as evidenced by the magnified views in Figures \ref{Figure5}b,c (in-plane and out-of-plane, respectively), the GLW behavior remains quite visible along the interface, in spite of the loss-induced attenuation. Figures \ref{Figure5}d,e shows two representative cuts (in-plane and out-of-plane, respectively), which are in very good agreement with the theoretical predictions. These latter are computed as for the eigenmode analysis in Figure \ref{Figure4}, but now assuming lossy constitutive parameters.

Overall, the above results confirm that GLW can be physically excited by dipole sources, and can be observed in the presence of small losses, together with the (isotropic and anisotropic) SWs naturally supported by the two halves.  

Although the  results above are limited to the case of hyperbolic anisotropy, GLWs can also be observed in junctions featuring elliptic-type anisotropy. Such scenario is discussed in the Supporting Information, paralleling the results in Figures \ref{Figure2}--\ref{Figure5}.
While most qualitative considerations still hold, there are some important differences that should be emphasized. In particular, GLW may exist for {\em any} rotation angle of the optical axis (with the exception of $\phi=0$ and $\phi=90^\circ$), and over spectral regions that are generally {\em open} intervals (i.e., for $k_z$ larger than a critical value). Moreover, beyond a critical rotation angle, additional GLW spectral regions may appear at lower wavenumbers. Finally, unlike the hyperbolic case, the choice of the isotropic half is now clear-cut, based on  the complementary character (i.e., capacitive for inductive-elliptic anisotropy, and vice versa).

%%%%%%%%%%%%%%%%%%%%%%%%%%%%%%%%%%%%%%%%%%
\section*{Conclusion}
%%%%%%%%%%%%%%%%%%%%%%%%%%%%%%%%%%%%%%%%%%

In summary, we have extended the concept of GW to flat-optics scenarios featuring planar junctions between isotropic and anisotropic metasurfaces. Specifically, we have demonstrated the existence of a new type of LW which exhibits the typical GW hallmarks, by combining the phase accumulation and evanescent decay, both in-plane (along the anisotropic surface) and out-of-plane (in the surrounding  medium). 

Our theoretical predictions are in excellent agreement with (eigenmode and dipole-excited) finite-element numerical simulations, also in the presence of losses, and shed  further light on the physics of GWs and LWs, two emerging  research areas that are still in their infancy. These outcomes may provide new opportunities for the nanoscale tailoring of light, via advanced control of surface polaritons, with potentially abundant applications ranging from nonlinear optics to optical sensing and imaging.

Current and future studies are aimed at exploring more in detail possible technological platforms for experimental observation of GLWs. 
Our results indicate that they may exist within broad parameter ranges, which are compatible with constitutive parameters exhibited by artificially engineered metasurfaces as well as natural low-dimensional materials. Clearly, realizing a planar isotropic-anisotropic junction (with precise control of the optical axis) constitutes a major technological challenge. One technologically feasible possibility could be nanopatterning two-dimensional materials such as graphene \cite{Gomez-Diaz:2016fo}.
As for natural anisotropic materials, thanks to the recent advances in the field of van der Waals junctions \cite{Zheng:2021ev,Yoon:2022ms}, this possibility may be within reach in the near future. It should also be mentioned that, while for simplicity we have limited our study to isotropic-anisotropic junctions, similar phenomena can occur interface between two anisotropic surfaces, with additional degrees of freedom for their control.
Within this framework, the extension of the analysis to elliptic dual (capacitive/inductive) configurations is relatively straightforward. However, in junctions involving two hyperbolic metasurfaces, the interpretation of the involved phenomena become more complex. 

Also of great interest is the study of resonant excitation and coupling mechanisms, as well as the possibility to attain exceptional points and other singularities by coupling two GLWs \cite{Debnath:2021gi,Khan:2022gs}. Finally, it would be interesting to explore to what extent some typical features of LWs (e.g., chiral coupling, unidirectional excitation, robustness to bend/curvature) are extended to GLWs.

\begin{acknowledgement}

 G. C. and V. G. acknowledge partial support from the University of Sannio via the FRA 2021 Program.
 A. A. has been partially supported by the Air Force Office of Scientific Research and the Simons Foundation.
 
\end{acknowledgement}

\begin{suppinfo}
Details on the analytical derivations and numerical simulations, and additional results can be found in the 
\href{https://www.dropbox.com/s/wb0dnqglns50714/G-LW_SI.pdf?dl=0m}{Supporting Information}.
\end{suppinfo}

\bibliography{G-LW_MT}

\end{document}